\documentstyle[preprint,aps]{revtex}

\begin{document}

\title{Symmetry transform in Faddeev-Jackiw quantization of dual models}

\author{C. P. Natividade,$^{1,2}$ A. de Souza Dutra$^2$ and H. Boschi-Filho$^3$}

\address{$^1$Instituto de F\'\i sica, Universidade Federal Fluminense\\Avenida Litor\^anea s/n, Boa Viagem, Niter\'oi, 24210-340 Rio de Janeiro, Brazil}

\address{$^2$Departamento de F\'\i sica e Qu\'\i mica, Universidade Estadual Paulista\\Avenida Ariberto Pereira da Cunha 333, Guaratinguet\'a, 12500-000 S\~ao Paulo, Brazil}

\address{$^3$Instituto de F\'\i sica, Universidade Federal do Rio de Janeiro\\Caixa Postal 68528, Rio de Janeiro, 21945-970 Rio de Janeiro, Brazil}

\maketitle

\begin{abstract}
We study the presence of symmetry transformations in the Faddeev-Jackiw
approach for constrained systems. Our analysis is based in the case of a
particle submitted to a particular potential which depends on an
arbitrary function. The method is implemented in a natural way and
symmetry generators are identified. These symmetries permit us to obtain
the absent elements of the sympletic matrix which complement the set of
Dirac brackets of such a theory. The study developed here is applied in
two different dual models. First, we discuss the case of a
two-dimensional oscillator interacting with an electromagnetic potential
described by a Chern-Simons term and second the Schwarz-Sen gauge
theory, in order to obtain the complete set of non-null Dirac brackets
and the correspondent Maxwell electromagnetic theory limit.
\end{abstract}

\bigskip
PACS number(s):

\newpage

\section{Introduction}
\label{sec:intro}
\setcounter{equation}{0}

Dual symmetries play a fundamental r\^ole in classical electromagnetic
theory as realized since the completion of its equations by Maxwell in
last century. In quantum theory however, these symmetries were not fully
appreciated until the works of Montonen and Olive \cite{mol77} and more
recently Seiberg and Witten \cite{swi94} in 3+1 dimensions and the study
of Chern-Simons (CS) theories \cite{C-S} in 2+1 dimensions. Since these
theories have gauge symmetries they are naturally constrained.

The study of constrained systems consist in a very interesting subject
which has been intensively explored by using different techniques
\cite{hte92}, alternatively to the pionneer Dirac's procedure
\cite{dir64}. In that original work, the constraints were classified
into two categories which have different physical meanings: first-class
constraints are related to gauge symmetries and second-class ones which
represent a reduction of the degrees of freedom. Besides its
applicability, the Dirac method present some difficulties when one
studies systems presenting only second-class constraints and there one
verifies the presence of symmetries despite of the gauge fixation. This
is what happens with 2D induced gravity where the $SL(2,R)$ symmetry was
not detected by convential methods but, analizing the anomaly equation
by Polyakov \cite{pol87,kpz88}. Later on, Barcelos-Neto \cite{bar94}
using the Dirac and sympletic methods reobtained this result and also
find a Virasoro hidden symmetry in the Polyakov 2D gravity.

From the canonical point of view, the study of symmetries can be
attacked with the Faddeev-Jackiw (FJ) sympletic procedure \cite{fja88}.
In this approach, the phase space is reduced in such a way that the
Lagrangian depends on the first-order velocities. The advantage of this
linearization is that the non-null Dirac brackets are the elements of
the sympletic matrix \cite{bwo92,mon93,wot95}. For gauge systems, this
matrix is singular and has no inverse unless a gauge fixing term is
included. The FJ method it is very simple to use nevertheless it does
not explicitly give all non-null Dirac brackets in the sympletic matrix.
Some of them are obtained only by use of the equations of motion
\cite{nbo96}. However, this problem may be circunvented if we consider
some symmetry transformations in the fields.

In this work we show how to implement this idea by using first in
section \ref{sec:symmetry} an example in one dimension where a particle
is submitted to an arbitrary potential which depends on a function which
will represent the constraints of the model. It is possible to verify
that the generators of the symmetries are given in terms of the
zero-modes of the sympletic matrix. Then we implement symmetry
transformations on the Lagrangian of the system so that new non-null
Dirac brackets emerge from the sympletic matrix. These ideas are
specially important to discuss dual theories where Dirac brackets
involving gauge fields are expected to appear. However, as we are going
to show in two different models these brackets do not come from a
canonical implementation of the sympletic method. We then show that
introducing convenient symmetry transformations, we can obtain the
complete set of Dirac brackets of the corresponding dual models. In
section \ref{sec:chern-simons} we apply this method to quantize the
problem of a charged oscillator in two space dimensions interacting with
an electromagnetic field described by a CS term. A similar system have
been investigated before using the Dirac method \cite{mat90} and it can
be undestood as an extesion of the quantum mechanical model of Dunne,
Jackiw and Trugenberger \cite{djt90}. Two of the present authors have
also investigated this system \cite{nbo96} using the FJ method but in a
noncanonical way, in the sense that we have not included a field to play
the r\^ole of the momentum of the CS field. This plannar system it is
also interesting to explore the r\^ole of the canonical quantization of
a particle under influency of a gauge field. Consequently, it can be
interpreted like a laboratory to dimensional reduction approach in other
more complicated models \cite{ana98}. In section
\ref{sec:schwarz-sen}, we explore the ideas introduced in section
\ref{sec:symmetry} to quantize, from the canonical poit of view, the
Schwarz-Sen model \cite{sse94}. The study of symmetry dualities reveal
to exist a conflict between electric-magnetic duality symmetries and
Lorentz invariance at the quantum level in the Maxwell theory
\cite{dte76,ggr97}. In a very interesting way, Schwarz and Sen proposed
a four dimensional action by using two gauge potentials, such that, the
duality symmetry is stablished in a local way. As a consequence, the
equivalence between this alternative theory and Maxwell's one is
demonstrated. Here, we use the features discussed in section
\ref{sec:symmetry} to obtain this equivalence. For our convinience, we
choose the Coulomb gauge in the treatments of both gauge theories
discussed on sections \ref{sec:chern-simons} and \ref{sec:schwarz-sen},
so that a parallel of the sympletic structure between that two different
dual models can be easily traced. Conclusions and final comments are
presented in section \ref{sec:comments}.

\section{Symmetry transform in the Faddeev-Jackiw approach}
\label{sec:symmetry}
\setcounter{equation}{0}

In order to show how the symmetry transformations are related to the
zero-modes of the sympletic matrix in the FJ approach we have made use
of a simple case, where a particle has been submitted to a potential
which depends on a constrained function. For a review on FJ method and
applications we refer to Ref. \cite{fja88,bwo92,mon93,wot95,nbo96}.

Let us start by considering the following Lagrangian
\begin{equation}\label{L0}
L^{(0)}=p_i {\dot q}_i + V (q,p,\Omega),
\end{equation}
where the potential is defined as
\begin{equation}
V(q,p,\Omega)=\lambda\Omega(q,p)-W(q,p),
\end{equation}
such that $\Omega(q,p)$ represent the constraints, $\lambda$ a Lagrange
multiplier and $W(q,p)$ the resultant potential. Following the steps of
the sympletic method we must build the sympletic matrix which contains
the Dirac brackets. Hence, we begin defining the matrix elements
\cite{fja88,bwo92,mon93,wot95,nbo96} 

\begin{equation}
\hat\rho\equiv(\rho_{ij})={\partial a_j\over\partial\xi_i}
-{\partial a_i\over\partial\xi_j},
\end{equation}
being $\xi_i\equiv(q_i,p_i)$ the generalized coordinates and $a_i$ the
coefficients of the velocities in the first-order Lagrangian $L^{(0)}$.
Therefore we have, by inspecting $L^{(0)}$ that $a^i_q\equiv p^i$ and
then
\begin{equation}
\rho^{ij}_{qp}=-{\partial a^i_q\over\partial p_j}
+{\partial a^j_p\over\partial q_i}=-\delta_{ij},
\end{equation}
since $a^j_p$ vanishes. Now, defining the vector
$\xi_i=(q_i,p_i,\lambda)$ and calculating the respective coefficients,
we obtain the matrix
\begin{equation}\label{rhohat0}
{\hat\rho}^{(0)}=
\left(
\begin{array}{ccc}
0 & -\delta_{ij} & 0 \\
\delta_{ij} & 0 & 0 \\
0 & 0 & 0 
\end{array}
\right)
\end{equation}
which is obviously singular, since $\det {\hat\rho}^{(0)}=0$. 
Then, in this case
we can not identify ${\hat\rho}^{(0)}$ as the sympletic matrix. This
feature reveals that the system under consideration is constrained~
\cite{bwo92,mon93,wot95}. A manner to circunvent this problem is to use
the constraints conveniently to change the coefficents $a^i(\xi)$ in the
first-order Lagrangian (\ref{L0}) and consequently obtain a rank-two
tensor which could be identified with the sympletic matrix. 

In the present case, we can build up an eigenvalue equation with the
matrix ${\hat\rho}^{(0)}$ and eigenvectors $v_i^{(0)}$ such that
\begin{equation}\label{zero-modes}
v_i^{(0)} {\rho^{(0)}}^{ij}=0.
\end{equation}
From the variational principle applied to Lagrangian (\ref{L0}) we find
the condition over the zero-modes

\begin{equation}\label{constraint0}
v_i^{(0)} \partial^i V^{(0)}\equiv\chi^{(0)},
\end{equation}
which generates the constraint $\chi^{(0)}$. If we impose that
$\chi^{(0)}$ does not evolve in time, we arrive at

\begin{equation}\label{notevolve}
{\dot\chi}^{(0)}=\left(\partial_i \chi^{(0)}\right) {\dot q}^i
\end{equation}
and since ${\dot\chi}^{(0)}$ is linear in ${\dot q}^i$ we can
incorporate this factor into Lagrangian (\ref{L0}). This operation means
to redefine the coefficients $a_i^{(0)}(\xi)$ in the form
\begin{equation}\label{lambda}
{\tilde a}_i^{(0)}(\xi)\to a_i^{(0)}(\xi)+\lambda\partial_i \chi^{(0)}
\end{equation}
where $\lambda$ is a Lagrange multiplier. Consequently the matrix
${\hat\rho}^{(0)}$ becomes
\begin{equation}
{(\tilde\rho)}_{ij}={\partial {\tilde a}_j\over\partial\xi_i}
-{\partial {\tilde a}_i\over\partial\xi_j}.
\end{equation}
After completing this, if $\det{(\tilde\rho)}_{ij}$ is still vanishing
we must repeat the above strategy until we find a nonsingular matrix. As
has been pointed out in the Refs. \cite{bwo92,mon93,wot95} for systems
which involve gauge fields it may occur that the matrix is singular and
the eigenvectors $v_i^{(m)}$ do not lead to any new constraints. In this
case, in order to obtain an invertible matrix, it is necessary to fix
some gauge. Such a case will be discussed in the following sections.

Going back to the Eq. (\ref{rhohat0}), we can see that the Eq. (\ref{notevolve}) is
satisfied for the eigenvector $v_i^{(0)}=(0,0,1)$. On the other hand,
from the Eq. (\ref{constraint0}) and the Lagrangian (\ref{L0}), we get
\begin{eqnarray}
{\chi}^{(0)} &=& v_i^{(0)} \partial^i V^{(0)}
= v_\lambda^{(0)} {\partial V^{(0)}\over \partial\lambda}=0 
\nonumber\\
&\equiv& \Omega(p,q),
\end{eqnarray}
so that $\Omega(p,q)$ is the primary constraint of the theory. In
order to include this constraint into the Lagrangian density we can use
a new Lagrange multiplier $\eta$ and make

\begin{eqnarray}\label{L1}
L^{(1)}&=&L^{(0)}\vert_{\Omega=0} +\dot\eta \Omega(q,p)
\nonumber\\
&=& p_i {\dot q}_i +\dot\eta \Omega(q,p) - W (q,p).
\end{eqnarray}
Hence, the new coefficients which contributes to the matrix are
$a^i_q=p_i$ and $a^i_\eta=\Omega$. Then, the iterated matrix 
${\hat\rho}^{(1)}$ reads
\begin{equation}\label{rho1}
{\hat\rho}^{(1)}=
\left(
\begin{array}{ccc}
0 & -\delta_{ij} & {\partial\Omega^j\over\partial q_i} \\
\delta_{ij} & 0 & {\partial\Omega^j\over\partial p_i}  \\
-{\partial\Omega^i\over\partial q_j}  
& -{\partial\Omega^i\over\partial p_j} & 0 
\end{array}
\right)
\end{equation}
which means that 
$\det{\hat\rho}^{(1)}\equiv\{ \Omega^i,\Omega^j\}_{PB}$.
Here, there are two possibilities. The first is when
$\det{\hat\rho}^{(1)}\ne 0$ and the matrix ${\hat\rho}^{(1)}$ is
invertible. The second one occurs when $\det{\hat\rho}^{(1)}= 0$. This
case is more interesting, since the eigenvectors
\begin{equation}
{v_i}^{(1)}=
\left(-{\partial\Omega\over\partial p_i}, 
{\partial\Omega\over\partial q_i}, 1  
\right)
\end{equation}
can be identified as the generators of infinitesimal
transformations. This feature will be quite explored in our analisys.

Going back to the matrix given by Eq. (\ref{rho1}), we notice the
absence of the diagonal elements. This is apparently natural since by
definition ${\hat\rho}^{(m)}$ is a rank-two tensor, and in general this
tensor is anti-symmetric. However, there are cases where the system
contains duality symmetry, as for example in the Chern-Simons theories
\cite{C-S,hte92}. In order to incorporate these elements into the
iterated matrix ${\hat\rho}^{(1)}$, we can suppose that some kind of symmetry
transform can be obtained from the zero-modes $v_i^{(m)}$.

Therefore, let us consider the following transformation in the auxiliary
coordinate\footnote{The transformation given by this equation has been
suggested in order to turn more simple the development of this section.
The final result obtained here can be checked via more general
transformations as well.} 

\begin{equation}\label{transf}
p_i\to p_i +f_i \qquad \Rightarrow \qquad 
\delta p_i= {\partial f_i\over\partial q_j}\; \delta q_j,
\end{equation}
being $f_i=f_i(q)$. Consequently, the modified Lagrangian 
${\tilde L}^{(0)}$ is given by

\begin{eqnarray}
{\tilde L}^{(0)} &=& (p_i +f_i) {\dot q}_i + {\tilde V} (q_i,p_i+f_i)
\nonumber\\
&=& (p_i +f_i) {\dot q}_i +\lambda{\tilde \Omega}(q_i,p_i+f_i)
-{\tilde W}(q_i,p_i+f_i)
\end{eqnarray}
and by implementing the sympletic method here we obtain the matrix

\begin{equation}
\left({\tilde\rho}_{ij}\right)^{(1)}=
\left(
\begin{array}{ccc}
f_{ij} & \delta_{ij} & {\partial\Omega^j\over\partial q_i} \\
-\delta_{ij} & 0 & {\partial\Omega^j\over\partial p_i}  \\
-{\partial\Omega^i\over\partial q_j}  
& -{\partial\Omega^i\over\partial p_j} & 0 
\end{array}
\right)
\end{equation}
in such a way that, according to Eqs. (\ref{zero-modes}) --
(\ref{lambda}) we arrive at

\begin{equation}
\det\left({\tilde\rho}_{ij}\right)^{(1)}=
\left\{{\tilde\Omega}^i,{\tilde\Omega}^j\right\},
\end{equation}
being $f_{ij}\equiv {\partial f_j\over\partial q_i}  
 - {\partial f_i\over\partial q_j}$.
Now, since $f_j$ is infinitesimal we can write
\begin{equation}
{\tilde\Omega}^i(q_j, p_j+f_j)
= {\tilde\Omega}^i(q_j, p_j)
+{\left({\partial{\tilde\Omega}^i\over\partial p_k}\right)}_{f=0}f^k,
\end{equation}
which implies that
\begin{eqnarray}
{\left\{{\tilde\Omega}^i,{\tilde\Omega}^j\right\}}_{PB}
&=& \left\{\Omega^i,\Omega^j\right\}
    \left[1+
    {\left({\partial{\tilde\Omega}^i\over\partial p_k}\right)}_{f=0}f^k
    \right]
\nonumber\\
&\equiv&0, \label{PB=0}
\end{eqnarray}
since $\left\{\Omega^i,\Omega^j\right\}=0$ has been considered here. The
above result reveals that the constraint algebra is preserved in front
of transformations (\ref{transf}).
Consequently, the matrix ${({\tilde\rho}^{(1)})}_{ij}$ remains
singular and the zero-modes in this case become
\begin{equation}\label{zero-modes1}
{v_i}^{(1)}=
\left(-{\partial{\tilde\Omega}^i\over\partial p_j}, 
{\partial{\tilde\Omega}^i\over\partial q_j}
-{\partial{\tilde\Omega}^i\over\partial p_k}f^k, 1  
\right),
\end{equation}
which implies that

\begin{equation}
{v_i}^{(1)} {\tilde\rho}_{ij}^{(1)}=
\left(0, 0, \{ {\tilde\Omega}^i,{\tilde\Omega}^j\}
\right),
\end{equation}
giving a null vector by virtue of Eq. (\ref{PB=0}). On the other hand,
the action of the zero-modes on the equations of motion yelds

\begin{eqnarray}
{v_i}^{(1)} \left[{\partial L^{(1)}\over\partial\xi_i}
- {d\over dt}\left({\partial L^{(1)}\over\partial{\dot\xi}_i}\right)
	    \right] =0
\nonumber \\
= {\dot\eta} \{{\tilde\Omega}^i,{\tilde\Omega}^j\}
- \{{\tilde\Omega}^i,{\tilde W}^j\},
\end{eqnarray}
in consequence of the Eqs. (\ref{L1}) and (\ref{zero-modes1}).
This means that no new constraints can arise from the equations of
motion. From the above equation we can get
\begin{equation}
 \{{\tilde\Omega}^i,{\tilde W}^j\}
= {v_i}^{(1)} \partial^i {\tilde W},
\end{equation}
and by virtue of the Eq. (\ref{PB=0}) we conclude that the zero-modes
are orthogonal to the gradient of the potential $\tilde W$, indicating
that they are generators of local displacements on the isopotential
surface. Consequently, they generate the infinitesimal transformations,
{\it i.~e.}, for some quantity $A(\xi)$ we must have
\begin{equation}\label{infinitesimal}
\delta A^\alpha=\left({\partial
A^\alpha\over\partial\xi_i}.v_i\right)\varepsilon,
\end{equation}
$\varepsilon$ being an infinitesimal parameter. 
From Eqs. (\ref{zero-modes1}) and (\ref{infinitesimal}) we have 

\begin{equation}
\delta q_i= - {\partial\tilde\Omega^i\over\partial p_l}\varepsilon_l,
\end{equation}
\begin{equation}
\delta p_i= \left( {\partial\tilde\Omega^i\over\partial q_l}
-{\partial\tilde\Omega^i\over\partial p_l} f^{il}\right)
\varepsilon_l,
\end{equation}
\begin{equation}
\delta \eta= \varepsilon,
\end{equation}
which permit us to show that the Lagrangian ${\tilde L}^{(1)}$ becomes 

\begin{eqnarray}
{\tilde L}^{(1)} &=& {L}^{(1)} + \delta {L}^{(1)}
\nonumber\\
 &=& {L}^{(1)} + {d\over dt}\left(p_i\delta q_i +
\delta\eta\tilde\Omega^i\right)\varepsilon,
\end{eqnarray}
which does not change the original equation of motion. Therefore, the
introduction of symmetry transforms into the original Lagrangian
$L^{(0)}$ leads to some elements of the sympletic matrix, which are the
Dirac brackets. Notice that this result has been obtained without lost
of the formal structure of the constraint algebra and the equations of
motion. In the following sections we present explicity examples in
field theory where these ideas will be explored in a quite way.


\section{Oscillator interacting with a Chern-Simons term}
\label{sec:chern-simons}
\setcounter{equation}{0}

Let us now consider the problem of charged particle subjected to a
harmonic oscillator potential moving in two dimensions and interacting
with an electromagnetic field described by a Chern-Simons term. This
problem was inspired in the Dunne, Jackiw and Trugenberger model
\cite{djt90} and has been considered before \cite{mat90,nbo96} in
different situations. Here we want to apply the canonical form of the
sympletic method which will lead us to the Dirac brackets but some of
them will be missing as we discussed in the previous section. Then we
use a convenient transformation to get the complete brackets set. So, we
start with the Lagrangian\footnote{Our
conventions here are: $\epsilon_{012}=\epsilon^{012}=1$ and
$g^{\mu\nu}={\rm diag}(-++)$.}

\begin{eqnarray}
L =&& {m\over 2}\left[\dot q_i(t)\dot q^i(t)
				-\omega^2 q_i(t)q^i(t)\right] 
-e\int d^2x  A_0(t,\vec x) \delta(\vec x -\vec q)
\nonumber\\
&&+e\int d^2x  A_i(t,\vec x) \delta(\vec x -\vec q)\dot q^i(t)
+\theta \int d^2x \epsilon_{\mu\nu\rho}A^\mu(t,\vec x)\partial^\nu A^\rho(t,\vec x),
\end{eqnarray}
where $q_i(t)$ is the particle coordinate with charge $-e$ on the plane
$(i=1,2)$, $A_\mu(t,\vec x)$ is the electromagnetic potential
$(\mu=0,1,2)$, and $\theta$ is the Chern-Simons parameter. 
In order to implement the sympletic method here we introduce the
auxiliary coordinate $p_i(t)$ through the transformation $\dot q^2 \to
2p\cdot q-p^2$ \cite{bwo92}, and define an auxiliary field $\Pi_i(t,\vec
x) = \epsilon_{ij} A^j(t,\vec x)$, so that we can write the above
Lagrangian as

\begin{equation}
{L}^{(0)} = \left[m p_i(t)- e A_i(t,\vec q)\right]\dot q^i(t) 
- \theta \int d^2x \Pi_i(t,\vec x)\dot A^i(t,\vec x) - V^{(0)}
\end{equation}
where $A_i(t,\vec q)=\int d^2x A_i(t,\vec x)\delta(\vec x-\vec q)$, 
and the potential is given by

\begin{equation}
{V}^{(0)} 
= {m\over 2}\left[p_i(t)p^i(t) + \omega^2 q_i(t)q^i(t) \right]
+ e A_0(t,\vec q) 
+2\theta \int d^2x \partial^i\Pi_i(t,\vec x) A_0(t,\vec x).
\end{equation}
Since this Lagrangian is linear on the velocities, we can identify the sympletic coefficients

\begin{equation}
a^{(0)}_{q_i(t)}= m p_i(t)- e A_i(t,\vec q),
\end{equation}
\begin{equation}
a^{(0)}_{A_i(t,\vec x)}= - \theta \Pi_i(t,\vec x),
\end{equation}
while the others are vanishing, which lead us to the matrix elements

\begin{equation}
\rho^{(0)}_{q_i p_j}= -m \delta_{ij}~,
\end{equation}
\begin{equation}
\rho^{(0)}_{q_i A_j}= e \delta_{ij} \delta(\vec y-\vec q)~,
\end{equation}
\begin{equation}
\rho^{(0)}_{A_i A_j}= 0~,
\end{equation}
\begin{equation}
\rho^{(0)}_{A_i \Pi_j}= \theta \delta_{ij} \delta(\vec x-\vec y).
\end{equation}
Defining the sympletic vector to be given by $y^\alpha=(\vec q,\vec p,
\vec A,\vec \Pi, A_0)$ we have the matrix

\begin{equation}
\rho^{(0)}_{\alpha\beta}= 
\left(
\begin{array}{ccccc}
0 & -m \delta_{ij} & e \delta_{ij} \delta(\vec y-\vec q) & 0 & 0 \\
m \delta_{ij} & 0 & 0 & 0 & 0 \\
-e \delta_{ij} \delta(\vec y-\vec q) & 0 & 0 
& \theta \delta_{ij} \delta(\vec x-\vec y) & 0 \\ 
0 & 0 & - \theta \delta_{ij} \delta(\vec x-\vec y) & 0 & 0 \\
0 & 0 & 0 & 0 & 0 
\end{array}
\right)
\end{equation}
which is obviously singular. The zero-modes come from the equation

\begin{equation}
{\partial V^{(0)}\over\partial A_0(t,\vec x)}= e \delta(\vec x-\vec q) +
2\theta\partial^i\Pi_i~,
\end{equation}
which implies the primary constraint $\chi^{(0)}= e \delta(\vec x-\vec
q) + 2\theta\partial^i\Pi_i$ (Gauss law). Using this constraint we can
built up the Lagrangian

\begin{equation}
{L}^{(1)} = {L}^{(0)} + \dot\lambda \chi^{(0)},
\end{equation}
where $\lambda$ is a Lagrange multiplier and now the potential reads

\begin{equation}
{V}^{(1)} = {{V}^{(0)}}\vert_{\chi^{(0)}=0}
= {m\over 2}\left[p_i(t)p^i(t) + \omega^2 q_i(t)q^i(t) \right].
\end{equation}
The new non-null velocity coefficient is given by
$a^{(1)}_{\lambda}= \chi^{(0)}= e \delta(\vec x-\vec q)+
2\theta\partial^i\Pi_i$, 
so that we have new matrix elements 
$\rho^{(1)}_{A_i \lambda}= 0$ and 
$\rho^{(1)}_{\Pi_j\lambda}=2\theta\partial_j\delta(\vec x-\vec y)$,
which lead us to the matrix $(y^\alpha=(\vec q,\vec p,
\vec A,\vec \Pi, \lambda))$

\begin{equation}
\rho^{(1)}_{\alpha\beta}= 
\left(
\begin{array}{ccccc}
0 & -m \delta_{ij} & e \delta_{ij} \delta(\vec y-\vec q) & 0 & 0 \\
m \delta_{ij} & 0 & 0 & 0 & 0 \\
-e \delta_{ij} \delta(\vec y-\vec q) & 0 & 0 
& \theta \delta_{ij} \delta(\vec x-\vec y) & 0 \\ 
0 & 0 & - \theta \delta_{ij} \delta(\vec x-\vec y) & 0 
& 2 \theta \partial_i \delta(\vec x-\vec y) \\
0 & 0 & 0 & - 2 \theta \partial_j \delta(\vec x-\vec y) & 0 
\end{array}
\right)
\end{equation}
which is still singular. The zero-modes will not lead to any new
constraints so we have to choose a gauge, which will be the Coulomb one
$(\vec\nabla \cdot \vec A=0)$ and we include it into the Lagrangian via
another Lagrange multiplier $\eta$

\begin{eqnarray}
{L}^{(2)} &=& {L}^{(1)} + \dot\eta\vec\nabla\cdot\vec A \nonumber\\
&=& {L}^{(0)} +\dot\lambda\chi^{(0)} +\dot\eta\vec\nabla\cdot\vec A
+{V}^{(1)} \nonumber\\
&=& \left[m p_i(t)- e A_i(t,\vec q)\right]\dot q^i(t) 
- \theta \int d^2x \Pi_i(t,\vec x)\dot A^i(t,\vec x) 
\nonumber\\
&&+\dot\lambda \left(e \delta(\vec x-\vec q) 
	+ 2\theta\partial^i\Pi_i\right)
+\dot\eta\vec\nabla\cdot\vec A
+{m\over 2}\left[p_i(t)p^i(t) + \omega^2 q_i(t)q^i(t) \right]
\end{eqnarray}
which implies the additional coefficient
$a^{(2)}_{\eta}= \vec\nabla \cdot \vec A$, 
and the new element 
$\rho^{(2)}_{A_i\eta}= \partial^i \delta(\vec x-\vec y)$. 
The sympletic tensor can then be identified with the matrix, 
$y^\alpha=(\vec q,\vec p, \vec A,\vec \Pi, \lambda, \eta)$, 

\begin{equation}
\rho^{(2)}_{\alpha\beta}= 
\left(
\begin{array}{cccccc}
0 & -m \delta_{ij} & e \delta_{ij} \delta(\vec y-\vec q) & 0 & 0 & 0\\
m \delta_{ij} & 0 & 0 & 0 & 0 & 0 \\
-e \delta_{ij} \delta(\vec y-\vec q) & 0 & 0 
& \theta \delta_{ij} \delta(\vec x-\vec y) & 0 & \partial_i \delta(\vec x-\vec y)\\ 
0 & 0 & - \theta \delta_{ij} \delta(\vec x-\vec y) & 0 
& 2 \theta \partial_i \delta(\vec x-\vec y) & 0 \\
0 & 0 & 0 & - 2 \theta \partial_j \delta(\vec x-\vec y) & 0 & 0 \\
0 & 0 & - \partial_j \delta(\vec x-\vec y) & 0 & 0 & 0
\end{array}
\right)
\end{equation}
which is not singular and can be inverted to give

\begin{eqnarray}
&&{\left(\rho^{(2)}\right)}^{\alpha\beta}=\nonumber\\
&&\left(
\begin{array}{cccccc}
0 & {1\over m} \delta_{ij} & 0 & 0 & 0 & 0\\
-{1\over m} \delta_{ij} & 0 & 0 
& -{e\over m\theta} D_{ij} \delta(\vec x-\vec q)& 0 
& -{e\over m}{\partial^i\over\nabla^2}\delta(\vec x-\vec q)\\
0 & 0 & 0 & -{1\over \theta} D_{ij} \delta(\vec x-\vec y)
& 0 & -{\partial^i \over\nabla^2}\delta(\vec x-\vec y)\\ 
0 &  {e\over m\theta} D_{ij} \delta(\vec x-\vec q)
& {1\over \theta} D_{ij}\delta(\vec x-\vec y)
& 0 &-{1\over 2\theta}{\partial^i\over\nabla^2}\delta(\vec x-\vec y)
& 0 \\
0 & 0 & 0 & {1\over 2\theta}{\partial^i\over\nabla^2}\delta(\vec x-\vec y) 
& 0 & -{1\over 2}{1\over\nabla^2} \delta(\vec x-\vec y) \\
0 & {e\over m}{\partial^j\over\nabla^2} \delta(\vec x-\vec q)
& {\partial^j \over\nabla^2} \delta(\vec x-\vec y)& 0 
& {1\over 2}{1\over\nabla^2} \delta(\vec x-\vec y)& 0
\end{array}
\right)\nonumber\\
&&
\end{eqnarray}
where $D_{ij}=\delta_{ij}-{\partial_i\partial_j\over\nabla^2}$. From
this result we can write down the following Dirac brackets of the theory:

\begin{equation}
\{q_i,p_j\} = {1\over m} \delta_{ij}~,
\end{equation}
\begin{equation}
\{p_i,\Pi_j\} = -{e\over m\theta} D_{ij} \delta(\vec x-\vec q)~,
\end{equation}
\begin{equation}
\{p_i,\eta\} = -{e\over m} {\partial^i\over\nabla^2} \delta(\vec x-
\vec q)~,
\end{equation}
\begin{equation}
\{A_i,\Pi_j\} = -{1\over\theta} D_{ij} \delta(\vec x-\vec y)~,
\end{equation}
\begin{equation}
\{A_i,\eta\} = -{\partial^i\over\nabla^2} \delta(\vec x-\vec y)~,
\end{equation}
\begin{equation}
\{\Pi_i,\lambda\} = - {1\over 2\theta} {\partial^i\over\nabla^2}
\delta(\vec x-\vec y)~,
\end{equation}
\begin{equation}
\{\lambda,\eta\} = -{1\over 2}{1\over\nabla^2} \delta(\vec x-\vec y)~.
\end{equation}
However, as was anticipated in section \ref{sec:symmetry}, 
some non-null  brackets are missing.
To overcome this situation let us consider the following transformation

\begin{equation}
\Pi_i(t,\vec x) \to \Pi_i^\prime(t,\vec x) 
- \epsilon_{ij} A^j(t,\vec x),
\end{equation}
on the Lagrangian of the system. So, following the same steps as shown
above we find a Lagrangian ${{L}^{(2)}}^\prime$ which is identical to 
${L}^{(2)}$ except for the substitutions

\begin{eqnarray*}
- \theta \int d^2x \Pi_i(t,\vec x)\dot A^i(t,\vec x) 
\to - \theta\int d^2x \Pi_i^\prime(t,\vec x)\dot A^i(t,\vec x) 
+ \theta\int d^2x \epsilon_{ij} A^j(t,\vec x) \dot A^i(t,\vec x) 
\end{eqnarray*}
and
\begin{eqnarray*}
2\theta\dot\lambda \partial^i\Pi_i (t,\vec x)\to 
2\theta\dot\lambda \partial^i\Pi_i^\prime (t,\vec x)
- 2\theta\dot\lambda \epsilon_{ij}\partial^i A^j(t,\vec x) 
\end{eqnarray*}
so that we find the coefficients

\begin{equation}
{a^{(2)}}^\prime_{A^i(t,\vec x)}= - \theta \Pi_i^\prime(t,\vec x)
+ \theta \epsilon_{ij} A^j(t,\vec x),
\end{equation}
\begin{equation}
{a^{(2)}}^\prime_{\lambda}= e\, \delta(\vec x-\vec q) 
+2\theta\partial^i\Pi_i^\prime(t,\vec x)
-2\theta\epsilon_{ij}\partial^i A^j(t,\vec x) 
\end{equation}
and the matrix elements

\begin{equation}
{\rho^{(2)}}^\prime_{A_i(t,\vec x)A_j(t,\vec y)}
= -2 \theta \epsilon_{ij} \delta(\vec x-\vec y)
\end{equation}
\begin{equation}
{\rho^{(2)}}^\prime_{A_i(t,\vec x)\lambda(t,\vec y)}
= 2 \theta \epsilon_{ij} \partial^j \delta(\vec x-\vec y)
\end{equation}

These new elements imply in another sympletic tensor which can also be
inverted. From this inverse we can reobtain the above Dirac brackets and
also others which were missing before in our treatment, 

\begin{equation}
\{A_i,A_j\} = -{1\over 2\theta} \epsilon_{ij} \delta(\vec x-\vec y)~,
\end{equation}
\begin{equation}
\{A_i,\lambda\} 
= {1\over 2\theta} \epsilon_{ij}\partial^j\nabla^{-2} 
\delta(\vec x-\vec y)~.
\end{equation}

At this point we can identify $\dot\lambda$ with $A_0$ so that we have
the quantized theory expressed in terms of the usual fields.


\section{Schwarz-Sen dual model}
\label{sec:schwarz-sen}
\setcounter{equation}{0}

Let us begin by introducing the basic idea of the Schwarz-Sen dual model \cite{sse94}. 
The proposal is to treat the problem of the conflict between eletric-magnetic duality and manisfest Lorentz invariance of the Maxwell theory. We mention, for instance, that by using the Hamiltonian formalism it can be shown that a non-local action emerges when one imposes the manifest Lorentz invariance and try to implement the duality symmetry \cite{dte76}. In order to circunvent this difficulty Schwarz and Sen proposed the introduction of one more gauge potential into the theory.

In this sense, the model is described by an action that contains two gauge potentials $A_\mu^a$ ($1\le a\le 2$ and $0\le \mu\le 3$) and is given by\footnote{Our conventions are: $\epsilon^{12}=1=-\epsilon^{21}$, and $1\le i,j,k\le 3$.}

\begin{equation}
S=-{1\over 2}\int d^4x \left( B^{a,i}\epsilon^{ab}E^{b,i} 
+ B^{a,i}B^{a,i}\right)
\end{equation}
being $E^{a,i}=-F^{a,0i}$, while $B^{a,i}=-(1/2)\epsilon^{ijk}F^{a}_{jk}$, and $F^{a}_{\mu\nu}=\partial_\mu A^{a}_\nu-\partial_\nu A^{a}_\mu$.
This action is separately invariant under local gauge transformations
\begin{equation}
\delta A^{a,0}=\psi^{a};\qquad
\epsilon A^{a,i}=-\partial^{i}\Lambda^{a}
\end{equation}
and duality transformations
\begin{equation}
A^{a,\mu} \to \epsilon_{ab} A^{b,\mu}.
\end{equation}
In terms of the gauge potentials, the corresponding Lagrangian density is given by

\begin{equation}\label{Ld}
{\cal L}=
{1\over 2} \epsilon^{ijk}(\partial_j A^{a}_k)
			\epsilon_{ab}(\dot A^{b}_i)
-{1\over 2}\epsilon^{ijk}(\partial_j A^{a}_k)
			\epsilon_{ab}(\partial_i A^{b}_0)
-{1\over 4} F^{a,jk}F^{a}_{jk}
\end{equation}
Now, the above Lagrangian density is of first-order in time derivative. In order to implement the sympletic method we can define an auxiliary field to turn more simple the subsequent calculations. Hence, let us consider the following field

\begin{eqnarray}
\Pi^{a,i} &=& \epsilon_{ab}\epsilon^{ijk}(\partial_j A^{b}_k)
\nonumber\\
&\equiv & \vec\Pi^{a}= \epsilon_{ab}\nabla\times \vec A^{b}
\end{eqnarray}
and the Lagrangian density (\ref{Ld}) becomes 

\begin{eqnarray}
{\cal L}^{(0)} &=&
{1\over 2} \vec\Pi^{a}\cdot\dot{\vec A^{a}}
-{1\over 2} \vec\Pi^{a}\cdot\nabla A^{a}_0
-{1\over 2} \vec\Pi^{a}\cdot\vec\Pi^{a}
\nonumber\\
&\equiv& {1\over 2} \vec\Pi^{a}\cdot\dot{\vec A^{a}}
- V^{(0)}\label{Ld0}
\end{eqnarray}
where $V^{(0)}= {1\over 2} \vec\Pi^{a}\cdot\nabla A^{a}_0
+{1\over 2} \vec\Pi^{a}\cdot\vec\Pi^{a}$. Therefore, the sympletic vector will be given as $\vec\xi^{(0)}=(\vec{A^{a}}, \vec{\Pi^{a}}, A^{a}_0)$. From the developments of section \ref{sec:symmetry}, it is easy to verify that the sympletic matrix correspondent to ${\cal L}^{(0)}$ is singular. The zero-mode vector in this case will be $v^{(0)}=(0, 0, v_{A_0}^{(0)})$ and the use of the Eqs. 
(\ref{zero-modes})-(\ref{notevolve}) will give origin to the constraint  

\begin{eqnarray}
\chi= \nabla\cdot  \vec\Pi^{a}=0
\end{eqnarray}
which can be incorporated into the new Lagrangian density via a Lagrange multiplier. Consequently, we have

\begin{eqnarray}
{\cal L}^{(1)} &=& {\cal L}^{(0)}\vert_{\chi=0}
+\dot\lambda \nabla\cdot  \vec\Pi^{a}
\nonumber\\
&=& 
{1\over 2} \vec\Pi^{a}\cdot\dot{\vec A^{a}}
+\dot\lambda \nabla\cdot  \vec\Pi^{a}
-{1\over 2} \vec\Pi^{a}\cdot\vec\Pi^{a},
\end{eqnarray}
which leads to another singular matrix

\begin{equation}\label{rho1ab}
{\left(\rho^{(1)}_{ab}\right)}_{ij}=
\left(
\begin{array}{ccc}
0 & -\delta_{ij} & 0 \\
\delta_{ij} & 0 & -\partial_i^x\\
0 & \partial_j^x & 0 
\end{array}
\right)\delta_{ab}\delta(\vec x-\vec y),
\end{equation}
where the sympletic vector has components $\xi^{(1)}=(\vec A^{a}, \vec\Pi^{a}, \lambda)$. On the other hand, the use of the eq. (\ref{constraint0}) implies that

\begin{eqnarray}
\vec {v}_{\vec {A}^{a}}- \nabla v_{\lambda}=0,
\end{eqnarray}
and no new constraints are generated. Here, we remark that the above relation means to derive the well-known gauge symmetry

\begin{equation}
\delta \vec {A}^{a}=\nabla\lambda;\qquad
\delta\vec\Pi^{a}=0;\qquad
\delta A_0^{a}=\dot\lambda.
\end{equation}
Hence, in this step it is necessary to impose a gauge fixing. If we adopt the Coulomb gauge, as was done in the previous section, the new Lagrangian density becomes

\begin{eqnarray}
{\cal L}^{(2)} &=& {\cal L}^{(1)}\vert_{\nabla\cdot\vec{A}^{a}=0}
+\dot\eta (\nabla\cdot\vec{A^{a}})
\nonumber\\
&=& 
{1\over 2} \vec\Pi^{a}\cdot\dot{\vec A^{a}}
+\dot\lambda (\nabla\cdot  \vec\Pi^{a})
+\dot\eta (\nabla\cdot\vec{A^{a}})
-{1\over 2} \vec\Pi^{a}\cdot\vec\Pi^{a},
\end{eqnarray}
and the sympletic matrix, given by

\begin{equation}\label{rho2ab}
{\left(\rho^{(2)}_{ab}\right)}_{ij}=
\left(
\begin{array}{cccc}
0 & -\delta_{ij} & 0 & -\partial_{i}^x \\
\delta_{ij} & 0 & -\partial_i^x & 0\\
0 & \partial_j^x & 0 & 0 \\
\partial_j^x & 0 & 0 & 0
\end{array}
\right)\delta_{ab}\delta(\vec x-\vec y),
\end{equation}
can be inverted to give

\begin{equation}\label{rho2ab-1}
{\left(\rho^{(2)}_{ab}\right)}^{-1}_{ij}=
\left(
\begin{array}{cccc}
0 & -\delta_{ab}D_{ij} & 0 & \partial_{i}^x \nabla^{-2}\\
\delta_{ab}D_{ij} & 0 & \partial_i^x\nabla^{-2} & 0\\
0 & -\partial_j^x \nabla^{-2}& 0 & \nabla^{-2}\\
-\partial_j^x \nabla^{-2}& 0 & -\nabla^{-2} & 0
\end{array}
\right)\delta(\vec x-\vec y),
\end{equation}
where $D_{ij}=\delta_{ij}+\partial_i^x\partial_j^x\nabla^{-2}$. The sympletic vector now is $\xi^{(2)}=(\vec{A^{a}}, \vec\Pi^{a}, \lambda, \eta)$, therefore from the above matrix we get

\begin{equation}\label{A,Pi}
\{ \vec {A^{a}}(\vec x), \vec\Pi^{b}(\vec y)\}_{D}
=-\delta_{ab}\left(\delta_{ij}
+{\partial_i^x\partial_j^x\over\nabla^{2}}\right)
\delta(\vec x-\vec y),
\end{equation}
which agrees with the result in Ref. \cite{ggr97} obtained from the Dirac procedure. It is important to notice that the matrix (\ref{rho2ab-1}) presents only one bracket, since by virtue of the dual symmetry it must contain diagonal elements like 
$\{\vec{A^{a}}(\vec x), \vec {A^{b}}(\vec y)\}_{D}$. This feature can be interpreted by considering the bracket (\ref{A,Pi}) as a dynamical one. The part of the sympletic matrix (\ref{rho2ab-1}) related with symmetries can not be identified directly. However, this term can be generated by means of a convenient symmetry transformation.

In order to implement this, let us consider the following transformation into the Lagrangian density (\ref{Ld0})

\begin{equation}\label{symtrans}
\vec\Pi^{a}\to \vec\Pi^{a\prime}
-\epsilon_{ab}\nabla\times \vec{A^{b}},
\end{equation}
so that we rewrite it as

\begin{eqnarray}
{{\cal L}^{(0)}}^\prime =
&+&{1\over 2}(\vec\Pi^{a\prime}-\epsilon_{ab}\nabla\times\vec{A^{b}}) 		\cdot(\dot{\vec A^{a}}-\nabla A^{a}_0 )
\nonumber\\
&-&{1\over 2}(\vec\Pi^{a\prime}-\epsilon_{ab}\nabla\times \vec{A^{b}})^2.
\end{eqnarray}
Now, by using the Eqs. (\ref{zero-modes}) and (\ref{constraint0}), it is easy to verify the presence of the constraint

\begin{equation}
\chi^\prime=\nabla\cdot\vec\Pi^{a\prime},
\end{equation}
and consequently

\begin{eqnarray}
{{\cal L}^{(1)}}^\prime = {{\cal L}^{(0)}}^\prime\vert_{\chi^\prime=0}
+\dot\alpha(\nabla\cdot\vec\Pi^{a\prime}).
\end{eqnarray}
The singular matrix corresponding to the above Lagrangian density is given by

\begin{equation}\label{G1ab}
{\left(G^{(1)}_{ab}\right)}_{ij}=
\left(
\begin{array}{ccc}
\epsilon_{ab}\epsilon_{ijk}\partial^k & 
-\delta_{ab}\delta_{ij} & 0 \\
\delta_{ab}\delta_{ij} & 0 & -\delta_{ab}\partial_i^x\\
0 & \delta_{ab}\partial_j^x & 0 
\end{array}
\right)\delta(\vec x-\vec y).
\end{equation}
Then, from the Eq. (\ref{constraint0}) we obtain the following zero-modes

\begin{eqnarray}\label{zero-modesa}
\vec {v}_{\vec{A}^{a}} = \nabla v_{\alpha^a};
\qquad 
\vec {v}_{\vec\Pi^{a}} = \nabla\times(\epsilon_{ab} \vec{v}_{\vec{A}^b}),
\end{eqnarray}
which confirm the two expected symmetries

\begin{equation}
\delta\vec{A^{a}}= \nabla\alpha^{a};
\qquad\qquad\qquad {\rm (gauge)}
\end{equation}
\begin{equation}
\delta\vec\Pi^{a\prime}=\nabla\times(\epsilon_{ab} A^{b});
\qquad\qquad {\rm (dual)}\label{dual}
\end{equation}
and no new constraints are generated. Therefore, we can adopt a gauge fixing. Choosing again the Coulomb gauge we arrive at

\begin{eqnarray}
{\cal L}_{GF} ={1\over 2}&&\left[ \;
(\dot{\vec A^{a}}-\nabla\dot\alpha^{a}-\vec\Pi^{a\prime} +\epsilon_{ab}\nabla\times\vec{A^{b}})\cdot\vec\Pi^{a\prime}
\right.\nonumber\\
&&\quad -\left.(\nabla\times \vec{A^{a}})^2
-(\dot{A^{a}}-\nabla\dot\alpha^{a}) \epsilon_{ab}\nabla\times\vec{A^{b}}
\right]_{\nabla\cdot\vec{A}^a=0}.
\end{eqnarray}
Identifying $\dot\alpha^{a}\equiv A_0^{a}$, we get 
$\vec{E^{a}}=-\dot{A^{a}}+\nabla A_0^{a}$, and consequently the gauge fixed Lagrangian density becomes

\begin{eqnarray}
{\cal L}_{GF}^\prime =&&
{1\over 2}(-{\vec E^{a}}-\vec\Pi^{a\prime} +\epsilon_{ab}\nabla\times\vec{A^{b}})\cdot\vec\Pi^{a\prime}
\nonumber\\
&& -{1\over 2}(\nabla\times \vec{A^{a}})^2
-\delta{\cal L}_{GF}\label{LGF}
\end{eqnarray}
and the gauge fixing term can be written as

\begin{eqnarray}
\delta{\cal L}_{GF} &=&
{1\over 2}({\vec E^{a}}+\vec\Pi^{a\prime}) 
\;\delta\vec\Pi^{a\prime}
\nonumber\\
&=&
{1\over 2}({\vec E^{a}}+\vec\Pi^{a\prime})
\;\delta(\epsilon_{ab}\nabla\times\vec{A^{b}})
\nonumber\\
&=&-{1\over 2}\nabla\times ({\vec E^{a}}+\vec\Pi^{a\prime})
\cdot\delta(\epsilon_{ab}\vec{A^{b}})
+{\rm surface\; terms}
\end{eqnarray}
where we used Eq. (\ref{dual}).

Before going on, it is important to make some remarks. First of all, we mention that from Eq. (\ref{LGF}) it is easy to infer that the Dirac bracket between the gauge fields in this case is given by

\begin{equation}
\{\vec{A^{a}}(\vec x), \vec{A^{b}}(\vec y)\}_D=
\epsilon_{ab}\nabla^{-2}\nabla\times\delta(\vec x-\vec y),
\end{equation}
which gives rise to a non-local commutation relation for the $\vec{A^{a}}$ field. This relation was obtained here within the context of the sympletic methodology starting from the use of the symmetry transform given by Eq. (\ref{symtrans}).

Another interesting feature is that from the use of Eq. (\ref{symtrans}) and of the gauge-fixed Lagrangian density (\ref{LGF}) we can show the equivalence between the Schwarz-Sen model and Maxwell theory. From (\ref{zero-modesa}), we notice that the variations over $\epsilon_{ab}\vec{A^{b}}$ leads to

\begin{equation}\label{constrainta}
\nabla\times(\vec\Pi^{a\prime}+\vec{E^{a}})=0
\end{equation}
and since at this stage the Gauss law can be used we conclude that

\begin{equation}
\delta\vec\Pi^{a\prime}=-\delta\vec{E^{a}}
=\nabla\times(\epsilon_{ab}\vec{A^{b}})
\end{equation}
and going back to Eq. (\ref{LGF'}) taking $a=1$, $b=2$, we have

\begin{eqnarray}\label{LGF'}
{\cal L}_{GF}^\prime \to {\cal L}_{M} =
{1\over 2}({\vec E^{1}}\cdot{\vec E^{1}}-
{\vec B^{1}}\cdot{\vec B^{1}})
\end{eqnarray}
which is the Maxwell Lagrangian density. From the Gauss law
$\nabla\cdot\vec\Pi^{a\prime}=0$ and the Eq. (\ref{constrainta}) we find that
the vector $\vec{u^{a}}=\vec\Pi^{a\prime}+\vec{E^{a}}\equiv 0$ implies that
$\nabla\cdot\vec{E^{a}}=0$. It is important to notice that the
$\vec\Pi^{a\prime}(\vec x)$ field is not the canonical momentum of the
electromagnetic theory but it represents here an auxiliary field in
order to implement the Faddeev-Jackiw method.


\section{Comments and conclusions}
\label{sec:comments}\setcounter{equation}{0}

In this work, we study the r\^ole of the symmetry transformations in 
the Faddeev-Jackiw approach. We verify that the generators of such a
transformation can be represented in terms of the zero-mode vectors of
the singular pre-sympletic matrix. Since the inverse of the sympletic
matrix contains elements which define the Dirac brackets of the
constrained system, it is natural to ask what happens when some brackets
do not appear in this inverse matrix. In our interpretation, these
elements are associated with some kind of symmetry transform which, on the
other hand are generated by zero-mode vectors. Hence, after a convenient
symmetry transformation we can complete the set of the fundamental
brackets of the models in question.

Here, we explore this strategy in three different situations. First, for
the case where a particle is submitted to a ``constrained potential''.
After we discuss the case of an oscillator in two space dimensions
coupled to a Chern-Simons gauge field and finally the Schwarz-Sen dual model for which our main goal was to obtain the corresponding Dirac brackets and how to describe its
equivalence with the Maxwell theory. In this point the zero-modes played
a very important r\^ole.

\acknowledgments
The authors acknowledge partial financial support from 
 Conselho Nacional de Desenvolvimento Cient\'\i fico e
Tecnol\'ogico, CNPq, Brazil.



\end{document}